\documentclass[twocolumn,showpacs,amsmath,amssymb,prl]{revtex4}

\usepackage{dcolumn}
\usepackage{bm}
\usepackage{graphicx}

\begin{document}

\title{Fermi Surface and Anisotropic Spin-Orbit Coupling of Sb(111) studied by Angle-Resolved Photoemission Spectroscopy}
\author{K. Sugawara,$^1$ T. Sato,$^{1,2}$ S. Souma,$^1$ T. Takahashi,$^{1,2}$ M. Arai,$^3$ and T. Sasaki$^3$}
\affiliation{$^1$ Department of Physics, Tohoku University, Sendai 980-8578, Japan}
\affiliation{$^2$CREST, Japan Science and Technology Agency (JST), Kawaguchi 332-0012, Japan}
\affiliation{$^3$ Computational Materials Science Center, NIMS, 1-2-1 Sengen, Tsukuba, Ibaraki 305-0047, Japan}

\date{\today}

\begin{abstract}	
High-resolution angle-resolved photoemission spectroscopy has been performed on Sb(111) to elucidate the origin of anomalous electronic properties in group-V semimetal surfaces.  The surface was found to be metallic despite the semimetallic character of bulk.  We clearly observed two surface-derived Fermi surfaces which are likely spin split, demonstrating that the spin-orbit interaction plays a dominant role in characterising the surface electronic states of group-V semimetals.  Universality/disimilarity of the electronic structure in Bi and Sb is discussed in relation to the granular superconductivity, electron-phonon coupling, and surface charge/spin density wave.

\end{abstract}


\maketitle

Bi- and Sb-based nanostructure materials have generated a considerable interest since they show variety of physical properties such as superconductivity in Bi clusters \cite{BiSuper}, semimetal-semiconductor transition and quantum size effect in thin films \cite{ThinFilm}.  These materials are also useful in devise application since their alloys and heterostructures are used for a highly efficient thermoelectric converter \cite{Thermo}.  These interesting properties owe to the characteristic electronic structure inherent to the semimetals, such as low carrier number and its high mobility.  To understand the origin of these intriguing behaviors, the electronic structure of Bi has been intensively studied \cite{AstPRLFS,AstPRBfull,AstPRLCDW,Bilambda,BiSOC}.  The band dispersion and the Fermi surface (FS) were determined by angle-resolved photoemission spectroscopy (ARPES) \cite{AstPRLFS,AstPRBfull,AstPRLCDW,Bilambda,BiSOC}, which revealed that the surface electronic structure is metallic in contrast to the semimetallic nature of bulk.  Two different surface-derived anisotropic FS sheets are reported in Bi(111), a small hexagonal electron pocket centered at the $\overline{\Gamma}$ point and six elongated hole pockets \cite{AstPRLFS}.   In spite of intensive ARPES studies on Bi surfaces, there is a considerable controversy on the nature of surface bands near {\it E}$_F$.  Ast and H$\ddot{\rm{o}}$chst reported the nested character of the hexagonal FS and observed the gap-opening at low temperatures and interpreted it in terms of the formation of the surface charge density wave (CDW) \cite{AstPRLCDW}.  On the other hand, Koroteev {\it et al}. concluded that the FS originates in the spin-split bands due to the spin-orbit coupling (SOC) \cite{BiSOC}, and suggested the formation of surface spin density wave (SDW).  These apparently contradicting conclusions strongly request the necessity for further investigation.  To clarify this issue is important not only in understanding the mechanism of anomalous physical properties in group-V semimetals, but also opens up the way to the application of new devises \cite{BiSOC}.  However, there is no firm experimental evidence so far to settle this controversy, since all the previous high-resolution ARPES studies were focused on Bi, but few on the other group-V semimetals \cite{SbARPESold,SbAst}.  Comparative study by Sb, which is a homologous element and has similar structural parameters to those of Bi \cite{SbARPESold,calc_TB,calc_LiuAllen}, would open a way to better and comprehensive understanding of the origin of the anomalous electronic properties of  Bi/Sb-based nanostructure materials.

In this Letter, we report high-resolution ARPES result of Sb(111).  We show that Sb(111) surface is metallic and possesses FSs qualitatively similar to those of Bi(111) \cite{AstPRLFS}, but remarkably different in the volume and symmetry.  Direct observation of the degeneracy of two surface bands exactly at the zone center establishes that the space-inversion symmetry breaks at the surface, and suggests that the bands are spin-split due to SOC.  We found no evidence for the gap opening at low temperature in contrast to the report on Bi(111) \cite{AstPRLCDW}.  All these results are well explained in terms of the anisotropic SOC, the weak electron-phonon ({\it e-p}) interaction, and the small effective mass of quasiparticle bands.

ARPES measurements were performed using a SES-2002 spectrometer with a high-flux discharge lamp and a toroidal grating monochromator.  We used the He I$\alpha$ (21.218 eV) resonance line to excite photoelectrons. The energy and angular resolutions were set at 3.5-12 meV and 0.2$^\circ$, respectively.  The angular resolution was 0.2$^\circ$.  A clean surface of sample was obtained by {\it in situ} cleaving in a vacuum of 2$\times$10$^{-11}$ Torr along the (111) plane.  Band calculation was performed by the linearized augmented plane wave method using the local density approximation \cite{citecalc} including the spin-orbit interaction.

\begin{figure}
\includegraphics[width=3.4in]{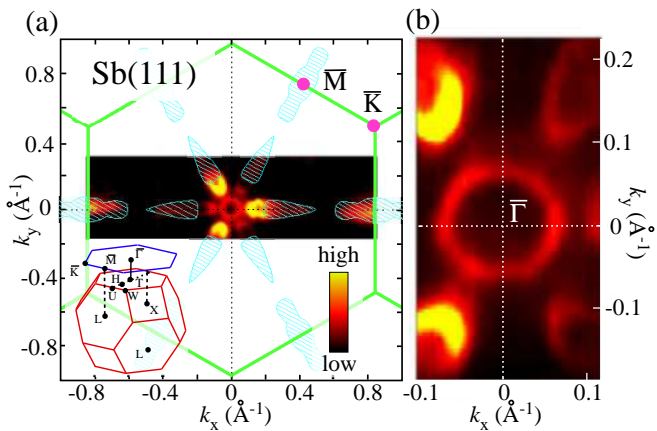}
\caption{(a) ARPES intensity plot at {\it E}$_F$ of Sb(111) as a function of two-dimensional wave vector, together with the calculated FS projected on the (111) plane.  ARPES intensity is integrated over the energy range of 20 meV centered at {\it E}$_F$.  Inset shows the bulk BZ and corresponding surface BZ.  (b) Expansion of (a) in the vicinity of the $\overline{\Gamma}$ point.}
 \end{figure}

Figure 1 shows the ARPES intensity plot at {\it E}$_F$ of Sb(111) as a function of two-dimensional (2D) wave vector.  We notice three different FSs, a small ring-like FS centered at the $\overline{\Gamma}$ point, six lobes with three-fold intensity variation, and an oval-shaped FS centered at the $\overline{\rm{M}}$ point.  The ring-like FS has no counterpart in the band calculation, suggesting the surface-derived character.  This demonstrates that the Sb(111) surface is metallic as Bi(111)\cite{AstPRLFS,AstPRBfull,AstPRLCDW,Bilambda,BiSOC}.  The observed strong intensity near the $\overline{\Gamma}$ point in the six elongated FSs is not expected from the band calculation, suggesting that this FS is also of surface origin.  The three-fold symmetry of spectral intensity indicates that the surface state is not simply confined within the surface bilayer with six-fold symmetry \cite{AstPRLFS}.  As for the six elongated FS away from the $\overline{\Gamma}$ point, the intensity shows an excellent agreement with the calculated FS which originates in the hole pocket at the {\it H} point of the bulk Brillouin zone (BZ) \cite{calc_TB,calc_LiuAllen}.  This feature is broad and subtle as compared to the ring-like FS, indicative of admixture of surface and bulk states which are essentially indistinguishable due to the surface resonance.  As seen in Fig. 1(a), the oval-shaped feature at the $\overline{\rm{M}}$ point is well reproduced by the projection of the calculated electron pocket centered at the {\it L} point \cite{calc_TB,calc_LiuAllen}.  As seen in Fig. 1, the small FS looks hexagonal-like rather than circle-like and possesses parallel regions indicative of a good nesting condition.  We estimated the Fermi momenta ({\it k}$_F$'s) by tracing the intensity maxima and found that the FS is six-fold in contrast to the trigonal-symmetric behavior of the elongated FSs.

\begin{figure}
\includegraphics[width=3.2in]{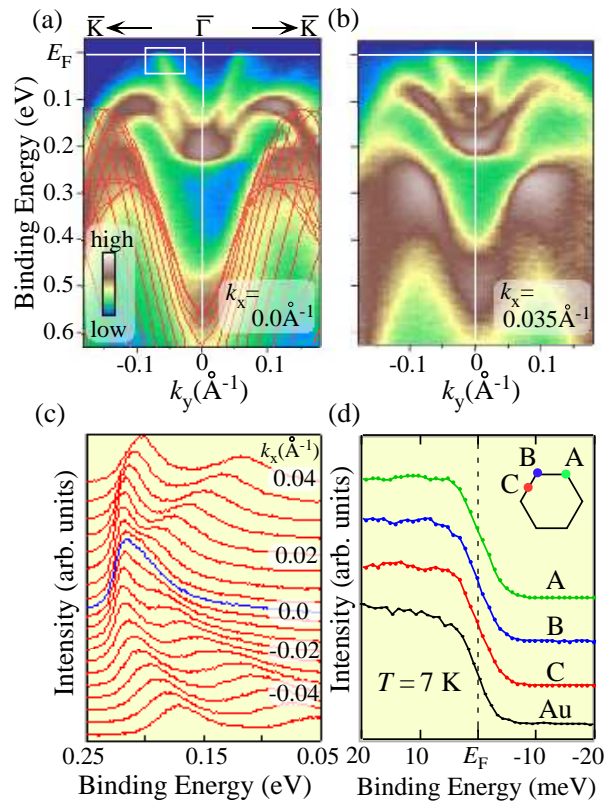}
\caption{ARPES spectral intensity plot of Sb(111) around the $\overline{\Gamma}$ as a function of  {\it k}$_{y}$ and binding energy for (a)  {\it k}$_{x}$ = 0.0 {\AA}$^{-1}$ and (b) {\it k}$_{x}$ = 0.035 {\AA}$^{-1}$.  Calculated bands along the $\overline{\Gamma}$-$\overline{\rm{K}}$ direction projected on the (111) plane are also shown by red curves in (a).  (c)  ARPES spectra measured around the zone center to show the degeneration of two bands. The spectrum at the zone center is indicated by a blue curve.  (d) Ultrahigh-resolution ($\Delta${\it E} = 3.5 meV) ARPES spectra in the close vicinity of {\it E}$_F$ at 7 K measured at three {\it k}$_F$ points (A, B, and C, in inset), compared with Au.  }
 \end{figure}

	In Fig. 2, we show ARPES intensity maps around the $\overline{\Gamma}$ point as a function of {\it k}$_y$ and binding energy for two different {\it k}$_x$ values.  At {\it k}$_x$ = 0.0 {\AA}$^{-1}$(Fig. 2(a)), we observe a highly-dispersive electronlike band with the bottom at $\sim$0.2 eV at $\overline{\Gamma}$ point.  Another band disperses in the energy region of 0.1 - 0.2 eV.  The former produces the small hexagonal-like electron pocket at the $\overline{\Gamma}$ point, while the latter produces the six elongated hole pockets.  These two bands degenerate at the zone center ($\overline{\Gamma}$ point).  We also find another relatively broad feature dispersing around 0.2 - 0.6 eV. This band overlaps with the projection of the calculated 5{\it p} band, demonstrating the bulk origin.  On the other hand, the two bands near {\it E}$_F$ which appear within the gap of projected bulk bands are assigned to the surface states. We found that the two surface bands degenerate only at the $\overline{\Gamma}$ point (Fig. 2(c)), and they are well separated at other {\it k}-points, as seen in the cut for {\it k}$_x$ = 0.035 {\AA}$^{-1}$  (Fig. 2(b)).  We now discuss the origin of these bands to resolve the controversy in the Bi surface \cite{AstPRLCDW, BiSOC}.  Important is whether the bands are spin-split or not.  In both bulk and surface, the time-reversal symmetry holds, requiring the constraint of spin-dependent energy dispersion, {\it E}({\it k}, $\uparrow$) = {\it E}(-{\it k}, $\downarrow$).  In addition, the space-inversion symmetry in the bulk requests that {\it E}({\it k}, $\uparrow$) = {\it E}(-{\it k}, $\uparrow$).  Combination of these requirements results in {\it E}({\it k}, $\uparrow$) = {\it E}({\it k}, $\downarrow$).  This does not lead to lifting of the spin degeneracy, hence the bands are doubly degenerate in the bulk.  However, on the surface where the space-inversion symmetry usually breaks, the time-reversal symmetry alone determines the character of bands.  Therefore, if the two bands are spin-split on the surface as in the case of Au(111) \cite{AuSOC, AuHuffner}, the band should show the degeneracy exactly at the $\overline{\Gamma}$ point because {\it E}(0, $\uparrow$) = {\it E}(0, $\downarrow$).  This behavior is indeed observed in the present ARPES experiment, suggesting that the two surface bands in Sb(111) are produced by SOC.  The strong anisotropy of the FSs / bands demonstrates that the spin-orbit interaction is quite anisotropic.  Clear observation of the degeneracy of the surface bands at the $\overline{\Gamma}$ point in Sb(111) is due to no overlapping between the surface and the bulk bands around the $\overline{\Gamma}$ point.  This is a great advantage of Sb(111), because the surface resonance, which significantly broadens and weakens the surface-state emission as observed in Bi(111), does not take place in Sb(111).  Moreover, similarity of the overall FS topology and band dispersion between Sb and Bi \cite{AstPRLFS}, in return, suggests that the two surface bands in Bi(111) are also caused by SOC.  To examine the possibility of gap opening on the electron pocket, we performed ultrahigh-resolution measurement at three representative {\it k}$_F$ points.  As shown in Fig. 2(d), the leading-edge midpoint at 7 K for all {\it k}$_F$ points coincides well with that of Au, demonstrating the absence of an energy gap.  This is in sharp contrast to the leading-edge shift (4-7.5 meV) for Bi(111) \cite{AstPRLCDW}.
		
\begin{figure}
\includegraphics[width=3.4in]{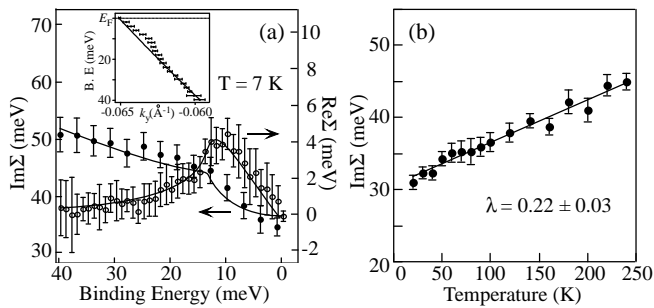}
\caption{(a) Imaginary (filled circles) and real (open circles) part of the self-energy at 7 K, Im$\Sigma$($\omega$) and Re$\Sigma$($\omega$), determined by the Lorentzian fitting of MDC on the electronlike band. Solid lines show the result of fitting by the bulk Debye model which satisfies a Kramers-Kronig relation.  The finite electron-electron scattering term is taken into account in Im$\Sigma$($\omega$). Inset shows the MDC peak position (solid circles) together with extrapolated parabolic bare-band dispersion (solid line).   (b) Temperature dependence of Im$\Sigma$(0).  Solid line represents the result of least-square fitting by linear function. $\lambda$ is the $e$-$p$ coupling constant.}
 \end{figure}
		
Figure 3 shows the result of numerical analysis on the electron band near {\it E}$_F$ indicated by a white rectangle in Fig. 2(a).  As seen in the inset to Fig. 3(a), the dispersion shows a weak kink at about 10 meV.  The estimated real and imaginary parts of self-energy, Re$\Sigma$($\omega$) and Im$\Sigma$($\omega$), are plotted in Fig. 3(a). Re$\Sigma$($\omega$) was obtained by assuming a parabolic bare band dispersion, and Im$\Sigma$($\omega$) was derived from the width of momentum distribution curves (MDC) as in the previous work \cite{Valla}.  Re$\Sigma$($\omega$) shows a maximum around 10 meV which corresponds to the sudden velocity change in the dispersion, and gradually goes to zero at the binding energy higher than 20 meV, reflecting the recovery of the energy dispersion to the bare band.  Im$\Sigma$($\omega$) keeps nearly constant but slightly decreases in the binding energy region of 40 - 15 meV, while it shows a relatively steep drop at lower binding energy.  The characteristic energy dependence in the self-energy indicates that surface electrons are coupled to a certain excitation, most likely phonon, since the Debye temperature of bulk Sb (211 K = 18 meV \cite{DebyeRef}) has a similar energy scale.  Figure 3(b) shows the temperature dependence of Im$\Sigma$(0), which exhibits almost linear behavior up to 250 K.  Estimated coupling constant $\lambda$ by using the approximation Im$\Sigma$(0;{\it T}) = $\pi\lambda${\it k}$_B${\it T} \cite{lambdaTemp} is 0.22 $\pm$ 0.03, which is classified into a weak-coupling regime.  This value is much smaller than that of Bi(111) which is as high as $\lambda$ = 0.7 \cite{AstPRBfull, BiSOC}.  It is noted that the estimated $\lambda$ for $\omega$ = 20 and 40 meV is 0.20 $\pm$ 0.03 in each, indicating that the coupling constant is independent of energy, at least within 40 meV.  This implies that the bulk density of states (DOS) does not show discernible modulation in the close vicinity of {\it E}$_F$ \cite{Bilambda}.  We fit the experimentally obtained self-energy by both the surface and the bulk Debye models with Debye energy ($\omega _D$) and $\lambda$ as free parameters, and found that the best fit by the surface model gives an anomalously high $\lambda$ (1.8), while the bulk model (solid lines) produces the value ($\lambda$ = 0.3) roughly consistent with the value estimated from the temperature dependence.  Since the bulk model also reproduces the $\omega _D$ (14 meV) close to the known bulk Debye energy (18 meV), these suggest that electrons at the surface are substantially coupled to the bulk phonon.

	We now discuss the surface states on Sb(111) in relation to the previous reports on Bi(111) \cite{AstPRLFS,AstPRBfull,AstPRLCDW,BiSOC}.  The hexagonal-like electron pocket and the six elongated hole pockets are commonly seen in both surfaces.  The main difference is that the {\it E}$_F$-intensity profile of the hole pockets in Sb is three-fold while that of Bi is six-fold.  This is understood in terms of the difference of the location of bulk valence-band maximum, which is at {\it T} point in Bi while it is between {\it T} and {\it W} points ({\it H} point) in Sb \cite{calc_TB,calc_LiuAllen,SbARPESold}.  When these bands are projected onto the surface BZ, a small FS is produced inside the surface-derived hexagonal-like FS in Bi \cite{BiSOC} while that is outside in Sb and overlaps with the surface-derived elongated hole pocket.  Since bulk bands have a three-fold symmetry, considerable overlapping between the surface and bulk bands would give rise to the observed  three-fold intensity variation for the six hole pockets in Sb(111).  We also find that the electron pocket is bigger in Sb.  Estimated {\it k}$_F$ points in Sb are 0.065 {\AA}$^{-1}$ and 0.071 {\AA}$^{-1}$ for $\overline{\Gamma}$-$\overline{\rm{K}}$ and $\overline{\Gamma}$-$\overline{\rm{M}}$ directions, respectively, while those of Bi are 0.053 {\AA}$^{-1}$ and 0.061 {\AA}$^{-1}$ \cite{AstPRLFS}.  The corresponding electron carrier number is 8.7$\times$10$^{12}$ cm$^{-2}$ for Sb and 5.5$\times$10$^{12}$ cm$^{-2}$ for Bi.  The difference in the volume is directly connected to the band width in an occupied side; bottom of the band at the $\overline{\Gamma}$ point is 0.22 eV in Sb (Fig. 2(a)) while that of Bi is 0.03 - 0.07 eV \cite{AstPRLFS,BiSOC}.  This leads to a fairly light electron mass of 0.09 $\pm$ 0.01{\it m}$_0$ in Sb, which is 2.5 times smaller than that of Bi \cite{AstPRLFS}.
	
The quantitative difference between Sb and Bi described above would be directly related to the strength of SOC.  In general, the magnitude of SOC can be estimated by the energy / momentum separation of two split bands as applied in Au(111) \cite{AuSOC,AuHuffner}.  In the case of Sb, it is strongly anisotropic and {\it k}-dependent, so that extraction of the accurate value is difficult.  However, the momentum separation of two FSs along the high symmetry line can be a good measure of it \cite{BiSOC}.  In this context, the narrower separation along the $\overline{\Gamma}$-$\overline{\rm{M}}$ direction of the electron and hole pockets in Sb (0.056 {\AA}$^{-1}$) compared to that in Bi (0.136 {\AA}$^{-1}$) \cite{AstPRLFS} would be a fingerprint of the weaker SOC in Sb.  This is supported by the experimental fact that the bottom of bands at $\overline{\Gamma}$ point is located at a higher binding energy in Sb than in Bi, which essentially tracks the trend of the reduced SOC in the band calculation of surface layers \cite{BiSOC}.  It is noted that the relatively weak SOC in Sb is reasonable since the atomic SOC in Sb is 2.5 times smaller than that of Bi \cite{calc_TB,calc_LiuAllen}.

	Next we discuss the consequence of {\it e-p} coupling. The weak {\it e-p} coupling in Sb(111) is naturally explained by the much lighter atomic mass of Sb (51) than Bi (83).  This would lead to a larger $\omega _D$, as a result a smaller $\lambda$.  In addition, the estimated partial DOS at {\it E}$_F$ for the electron pocket \cite{DOS} in Sb(111) is 2 to 5.5 times as small as that of Bi(111), so that it would also reduce $\lambda$.  We think that the difference in $\lambda$ between Bi and Sb explains the appearance of superconductivity in granular Bi \cite{BiSuper} but not in Sb.  Indeed, the estimated superconducting transition temperature by McMillan formula, {\it k}$_B${\it T}$_c$ = $\hbar$$\omega _D$ exp (-1/$\lambda$) \cite{lambdaTemp}, is {\it T}$_c$ = 1 K in Sb, which is significantly smaller than that determined in Bi (8 K) \cite{AstPRBfull}, implying the considerable difference in the superconducting behavior of granular system.
	
We think that occurrence of CDW in Sb(111) is quite unlikely, because we did not observe an energy gap down to 7 K or a superlattice spot in the LEED pattern.  In addition, the excitation process which connects two nested regions has to involve a spin flip, since the surface bands are spin-split with opposite spins for opposite {\it k} vectors.  This spin-flip process is also unfavorable to the CDW formation.  The small value of the nesting vector {\it Q} = 0.13 {\AA}$^{-1}$, corresponding to the long-range real-space periodicity of about 25 {\AA}, may not be a good condition for the CDW formation.  Then, a question arises in why the energy gap is seen in Bi(111) \cite{AstPRLCDW} but not in Sb(111).  The weaker {\it e-p} coupling in Sb compared with that of Bi \cite{AstPRBfull, Bilambda} would remarkably reduce the CDW transition temperature ({\it T}$_{\rm{CDW}}$).  In addition, the lower partial DOS at {\it E}$_F$ of the electron pocket \cite{DOS} in Sb than that of Bi is also unfavorable to the high {\it T}$_{\rm{CDW}}$.  The three-fold symmetry of six elongated hole pockets in Sb as seen in Fig. 1(a) unlike the overall six-fold symmetry in Bi may degrade the CDW stability on the electron pocket by the scattering between the hole and electron FSs.  We note here that although formation of the surface CDW is quite unlikely, the possibility of surface SDW is not excluded due to the spin-split nature of surface bands.  To elucidate this point, further detailed ultrahigh-resolution ARPES measurement combined with spin-resolved experiments at various locations on FSs is highly desired.

In conclusion, we demonstrated that the anisotropic spin-orbit interaction characterizes the surface electronic structure of Sb.  The electron-phonon and the spin-orbit coupling are found to be the essential factors in understanding the anomalous physical properties of Bi- and Sb-surfaces.  Those couplings are remarkably weak in Sb(111) as compared to those of Bi(111).  These experimental results together with the smaller effective mass of quasiparticle band in Sb(111) consistently explain why the CDW or SDW energy gap and the granular superconductivity are absent in Sb in contrast to Bi.

This work is supported by grants from MEXT and CREST-JST of Japan.  S.S. thanks JSPS for financial support.

\end{document}